# Multi-hop Energy-efficient Control for Heterogeneous Wireless Sensor Networks Using Fuzzy Logic

Alexandre M. Melo Silva, Christiano C. Maciel, Suelene do Carmo Corrêa
Federal University of Pará, UFPA
Belém, Brazil
amelo, christiano, suelene{@ufpa.br}

*Abstract* — **Wireless Sensor Networks (WSN) have severe energy constraints imposed by limited capacity of the internal battery of sensor nodes. These restrictions stimulate the development of energy-efficient strategies aimed at increasing the period of stability and lifetime of these networks. In this paper, we propose a centralized control to elect more appropriate Cluster Heads, assuming three levels of heterogeneity and multi-hop communication between Cluster Heads. The centralized control uses the k-means algorithm, responsible for the division of clusters and Fuzzy Logic to elect the Cluster Head and selecting the best route of communication. The study results indicate that the proposed approach can increase the period of stability and lifetime in WSN.**

## I. Introduction

Representing a subclass of wireless ad hoc networks, the Wireless Sensor Networks (WSN) are considered as a new generation of real-time embedded systems with computational resources, energy and limited memory. These networks may have large numbers of wireless sensor nodes, introduced in a region of interest, cooperating with each other to accomplish substantial tasks as reporting information about a monitored phenomenon to a collection point, called Base Station (BS).

One of the main barriers presented in these networks is related to energy restriction due to the limited capacity of the internal batteries of the sensor nodes. Several factors are culminating to battery drain of the nodes, being the radio module the major consumer of energy in the sensor nodes during the data transmission process [1].

The Energy consumption can be reduced, assuming that some nodes may only send data to the BS. Hierarchical WSN organize their nodes in clusters and elect a group leader node, called cluster head (CH). The CH is responsible for collecting data from all nodes in its cluster, can aggregate and then forwards them to the BS [2].

The hierarchical structure can be formed by two types of networks, homogeneous or heterogeneous. In homogeneous networks, the nodes have the same characteristics, such as energetic capacity, processing and radio [3]. In heterogeneous structures, some sensor nodes may have different hardware requirements, such as better energetic capacity. These sensor nodes provide to the network a longer period of stability [4, 5].

Studies that consider the heterogeneity of nodes can be found in [3,4,5,6,7,8].

The algorithm LEACH (Low Energy Adaptive Clustering Hierarchy) proposed by [9] is the most popular solution found in the literature for clustering and serves as the basis for numerous studies focused on clustering. As in LEACH, the major problem of these algorithms is to use local information based on probability calculations for the election of the leaders of the clusters. This type of selection may generate CHs very close to the edge of the network, increasing the power dissipation due to the transmission distance of the nodes to the CH. Another major problem with the form of election algorithm used by LEACH, is the lack of discriminatory treatment of the energetic discrepancies of nodes that form the network, since the selected CHs must have sufficient energy resources to support the transmission loads of the nodes associated to it.

Considering the obstacles related to the election of CHs based on local information without considering placement criteria, this paper proposes a strategy for the election of the ideal CH for heterogeneous WSNs using fuzzy logic based on BS centralized information. This centralized control sets the CH based on information acquired during the network formation. The information collected is used to load the k-means algorithm, responsible for the division of clusters, and the fuzzy system, which is responsible for selecting the leader of each group formed by the k-means algorithm.

The fuzzy system is also responsible for determining which CHs go through the process of transferring data of the farthest CH; the elected nodes that exceed the communication threshold with the BS should send their data to the closest nodes. For this selection the Fuzzy system uses as criteria: communication distance and energy levels. The criteria used for selection of CHs are: energy level, centrality and proximity to the BS. The insertion of three levels of heterogeneity, the use of BS centralized information and communication across multiple hops allow electing CHs well positioned and with adequate energy levels to sustain the transmission load of the cluster, increasing the period of network stability and lifetime.

The remainder of the paper is organized as follows: Section II presents some works related to the election of CHs, examples of motivation to the proposed algorithms and definitions of terms. Section III provides a brief introduction to fuzzy logic and k-means algorithm. The analytical models and the modeling used in the proposal is presented in Section IV.

Section V discusses the results and the simulation for the fuzzy system. Finally, Section VI concludes the paper.

## II. RELATED WORKS

The algorithm LEACH (Low Energy Adaptive Clustering Hierarchy), proposed by [9] elects the CH based on local information at each new cycle (round). To electing the node the number chosen should be less than the threshold *T*. After the election, the new CH sends announcement messages to all the network nodes. The remaining members decided to which CH they should connect.

Although the LEACH presents a hierarchical structure that can reduce the energy consumption in WSN, it doesn't take any position criterion during the election of the CH, and may select CH near the edge of the network. Another problem of the algorithm is described in [8] SEP, proving that the LEACH is not efficient in heterogeneous structures by not considering the energy discrepancy of the network nodes.

The DEEC algorithm proposed by [4] uses local information to the CH election and is able to treat heterogeneous network. The CHs are selected using a probability based on the residual energy of nodes and average energy of the network. The heterogeneous inserted in DEEC concerns only to the distinguished energy capacity of a set of nodes of the network. The images presented in [4] clearly illustrate the election of leaders close to the network edge.

Based on the DEEC, the E-DEEC algorithms proposed by [5] and LEACH-HPR proposed by [3] also considers the residual energy of the nodes in the network, and uses local information to the CH election process. The differential of the algorithms is the insertion of three types of nodes with different energy levels: normal nodes, advanced nodes and super nodes. Such algorithms allow the authors to prolong the fully functional network period with the addition of the super nodes. To minimize power dissipation during communication between the BS and the farthest CHs, the authors of the LEACH-HPR propose a multi-hop algorithm between the elected CHs, similar to ACHTLEACH proposed in [10].

In [11] it is proposed an algorithm that considers a hybrid mobile strategy in order to distribute the energy consumption throughout the network, with the CH moving to a place of higher energy concentration when an event occurs and with the possibility of controlling its transmission power. However, this displacement can maximize the energy spent in sending data to the BS, the hybrid mobile strategy of the BS-CH algorithm allows the movement of the base station to minimize the distance to the CH and the consumption in data transmission, disregarding the energy consumed, assuming that the base station is mobile, and without information if the power supply is direct.

## III. FUZZY LOGIC AND K-MEANS ALGORITHM

Proposed by [12] the fuzzy logic uses methods in order to control the vague language and imprecision used by humans, through a set of values represented by linguistic variables. Each set has a range of values directly associated with semantic rules. Unlike the Boolean logic, in fuzzy logic, the evaluation of a given proposition can comprehend values, degrees of relevance, varying in the range of 0 to 1.

The k-means algorithm is a data grouping technique very popular for its ease implementation. Normally the clustering algorithms are widely used in applications that require generation of patterns, dividing objects into groups useful or meaningful [13]. In the proposal presented in this paper the k-means algorithm is loaded with the coordinates of all the nodes of the network. Thus the algorithm generates a pattern by dividing the clusters that are formed by the closest nodes.

## IV. THE SOLUTION MODELING

For the solution proposed in this paper, the choice of cluster head occurs at every round. A round ends at the end of the aggregation process and data transmission to the BS. Basically the process is divided into three steps: (i) The first step is the beginning of the clustering process. The division of clusters is done using k-means algorithm, (ii) the second step is divided into two phases. The first phase consists of selecting the CH for each cluster formed by k-means. The data of each cluster are loaded on Fuzzy System, and this, based on the adopted criteria selects the most appropriate CHs for each cluster. In the second phase, after the results of the leaders selecting, the BS calculates the distance of the CHs that exceeded the communication threshold for all elected Cluster Heads. This calculation is used to determine to which leader the farthest CH must transmit their already aggregated data. Upon completion of the two phases, the BS sends broadcast messages to the network nodes informing the ID of the leader of the group so that the nodes can send data to their respective CH, (iii) The third step concerns the data aggregation process by the CH. This process is to compress the data and send them to BS.

In the nodes association process of a particular cluster to the respective CH, the nodes that form the cluster receive the BS message informing to which leader they should send membership requests. The leader reserves a TDMA slot to each node associated, after, they can transmit their data. The leaders responsible for propagating the already aggregated data of the farthest nodes receive along with the BS announcement message, a TDMA slot reservation request for the propagation process.

### A. Energy dissipation model

The energy dissipation model adopted is similar to the model used by LEACH and consists of the dissipated energy during the transmission and reception of k-bit message at a distance d. Tthe equations of consumption are (1) and (2).

$$E_{Tx}(k,d) = E_{elec} * k + \varepsilon_{fs} * k * d^2 \quad for\ d < d_0$$
$$E_{Tx}(k,d) = E_{elec} * k + \varepsilon_{amp} * k * d^4 \quad for\ d \geq d_0 \quad (1)$$

$$E_{Rx}(k,d) = E_{elec} * k \qquad (2)$$

The dissipated energy during the transmission and reception of radio is represented by $E_{elec} = 50\ nJ/bit$. For the transmission amplifier can reach an acceptable level, two models are used, **Free space** or **Multipath** depending on the distance between transmitter and receiver. If this distance does not exceed the threshold $d_0$, the Free space is used. The energy dissipated by the transmission amplifier is given by $E_{fs} = 10\ pJ/bit/m^2$, If the threshold is exceeded the model used is Multipath $E_{amp} = 0.0013\ pJ/bit/m^4$. where $d_0 = \sqrt{E_{fs}/E_{amp}}$ [5,9].

The Multipath generates more energy dissipation in the communication process. Energy-efficient strategies must consider the placement of the nodes in the CHs selection process in order to maintain the use of the Free Space.

*B. CH election criteria*

The criteria used to load the fuzzy system are: (i) Energy – The energy level of each network is represented by the linguistic variable Battery and has the linguistic values, Low, Moderate and High. Considering that the network has heterogeneous characteristics concerning to the nodes energy, the super nodes and advanced nodes represent greater chance to election of the CH. Fig. 1.

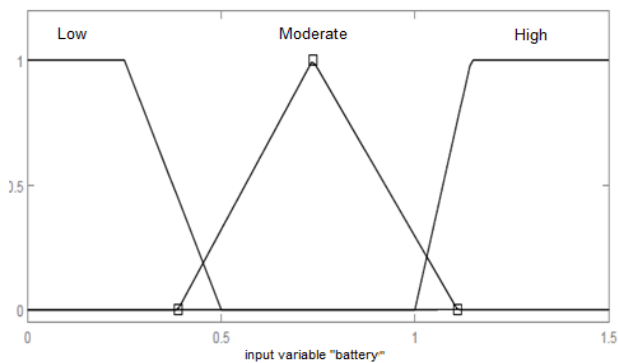

Figure1. Linguistic variable Battery.

(ii) Centrality – The variable centrality refers to the node placement in relation to the CH. The linguistic values that represent the variable centrality are: Near, Moderate and Far.

The lower the value of centrality, the nearest the node is from the cluster center, Fig. 2.

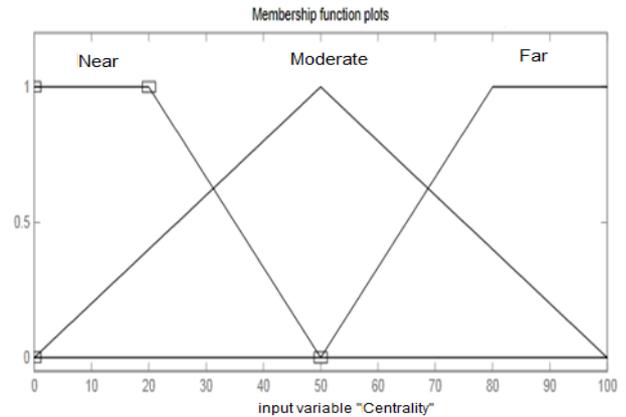

Figure 2. Linguistic variable Centrality.

(iii) Distance to BS – This is represented in the system as DistBS, their linguistic values are: Near, Moderate and Far. The criterion distance to the BS is used to generate an approach of the CH with the BS, aiming to minimize the energy consumption of the CH during the data transmission phase, Fig. 3.

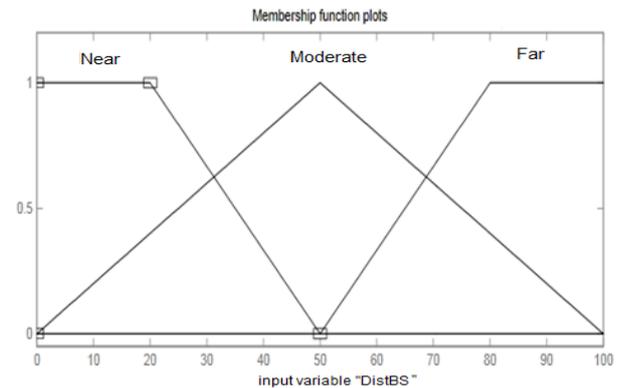

Figure 3. Linguistic variable Distance to BS.

The variables described above with its degrees of membership, correspond to the inputs of the fuzzy system.

The consequents of the system are the output fuzzy sets that will determine the optimal CH in a cluster *x*, and have the following linguistic values: Very weak, Weak, Medium, Strong and Very strong, Fig. 4.

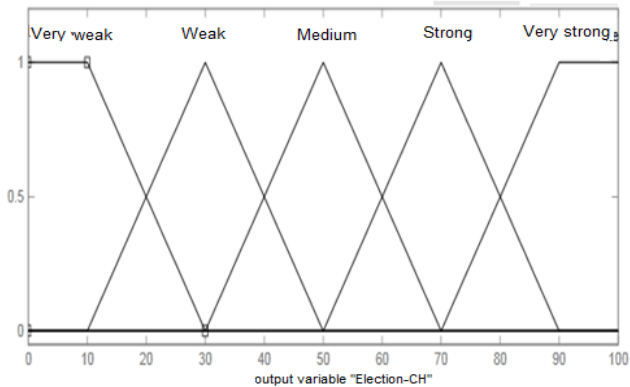

Figure 4. Outputs of the fuzzy system, consequents of the system.

In the mapping process of the output fuzzy sets to crisp values (defuzzification), each sensor node presents its respective output value of the fuzzy system. The largest output values indicate the cluster heads selected for the current round.

The Rule Base of the system contains twenty-seven rules, $3^3=27$, Table I. The better condition to the cluster head election is given by the following rule: *if Centrality is Near and Battery is High and DistBS is Near then Election-CH or Output is Very strong*.

### C. Nodes centrality and distance to BS

It is assumed that the BS knows the energy level and placement of the nodes, information that is sent at the beginning of the network formation and the nodes dissipate energy in this process. To determine the centrality values, the BS selects each node and calculates the Euclidean Distance of the nodes to the center of their respective clusters defined by the k-means algorithm in (3). The node with higher centrality (CH), will allow more smoothly power dissipation during communication comparing to other associated nodes.

$$d(P_i, P_j) = \sqrt{(x_i - x_j)^2 + (y_i - y_j)^2} \qquad (3)$$

TABELA I RULE BASE OF THE FUZZY SYSTEM

| Rules | Inputs | | | Output |
|---|---|---|---|---|
| | *Centrality* | *Battery* | *DistBS* | *Election-CH* |
| 1 | Far | High | Far | Weak |
| 2 | Far | High | Moderate | Weak |
| 3 | Far | High | Near | Medium |
| 4 | Far | Low | Far | Very weak |
| 5 | Far | Low | Moderate | Very weak |
| 6 | Far | Low | Near | Very weak |
| 7 | Far | Moderate | Far | Very weak |
| 8. | Far | Moderate | Moderate | Weak |
| 9 | Far | Moderate | Near | Very weak |
| 10 | Moderate | High | Far | Medium |
| 11 | Moderate | High | Moderate | Strong |
| 12 | Moderate | High | Near | Strong |
| 13 | Moderate | Low | Far | Very weak |
| 14 | Moderate | Low | Moderate | Weak |
| 15 | Moderate | Low | Near | Very weak |
| 16 | Moderate | Moderate | Far | Weak |
| 17 | Moderate | Moderate | Moderate | Strong |
| 18. | Moderate | Moderate | Near | Medium |
| 19 | Near | High | Near | Very strong |
| 20 | Near | High | Far | Strong |
| 21 | Near | High | Moderate | Very strong |
| 22 | Near | Low | Far | Very weak |
| 23 | Near | Low | Moderate | Very weak |
| 24 | Near | Low | Near | Very weak |
| 25 | Near | Moderate | Far | Medium |
| 26 | Near | Moderate | Moderate | Strong |
| 27 | Near | Moderate | Near | Strong |

The same process occurs in the distance calculation of each node to the BS. However, it is worth noting that this criterion is very important when the BS is not placed too far of the cluster at issue. Thus, the arrangement of the network must be considered for the distance criterion remains valid. With this, secure rules are defined to not generate CHs near the network edge, since the centrality is more important.

### D. Multi-hops between Cluster Heads

The multi-hop strategy is used for CHs that exceed the communication threshold, $d > d_0$, to BS. The approximation to the BS, presented previously, does not always apply, since, depending on the layout of the nodes, the clusters may be formed too far. The multi-hop strategy is implemented to minimize de energy dissipation during the communication. In this proposal, for determining which leaders will pass through the data propagation process of the farthest CHs, the fuzzy system adopts distance and energy level criteria. First is used the calculation (3) to determine the communication distances between the farthest CHs and the rest of the elected leaders. The fuzzy system is loaded with the result of the energy levels and distance calculation of each leader. The linguistic variables that represent the battery level of the leaders and the distance of all elected CHs to the farthest CHs are respectively presented by Fig. 5 and Fig. 6.

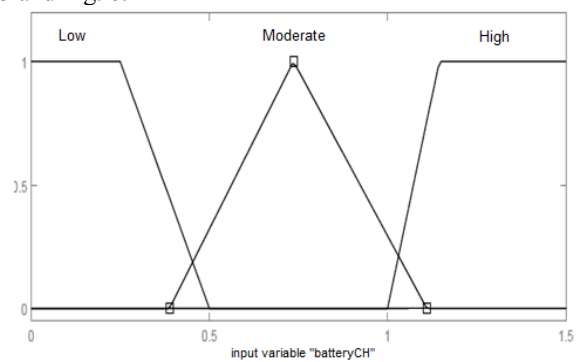

Figure 5. Variable BatteryCH

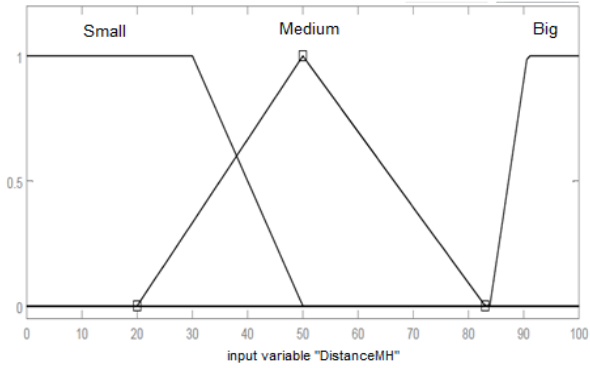

Figure 6. Distance to multiple hops

Along the simulation time it is natural that the nodes have a decline in energy level. Thus, to avoid that leader nodes with low energy level pass through the transferring data process of the farthest nodes, the energy criterion is used in a discriminatory manner to exclude these nodes of the process. The super and advanced nodes have more chance to participate of the propagation process, since its energy resources are increased.

The variable DistanceMH é is used to select the leader nodes that will propagate the already aggregated data of the CHs whose communication distance to the BS exceed the threshold $d_0$. The linguistic values of the variable are: Small, Medium and Big, the universe of the discourse ranges between 0 and 100 m. The membership function representing the fuzzy set Big is defined from 83 m, the established communication threshold based on the quotient of the signal power amplifiers, $d_0 = \sqrt{E_{fs}/E_{amp}}$ .

The rule base has nine rules, due the existence of two antecedents as inputs, $3^2$. The better condition to the selection of the optimal propagation CH is given for: *If DistanceMH is Small and BatteryCH is High then Output is Strong*. The Fig. 7 shows the system surface graphic.

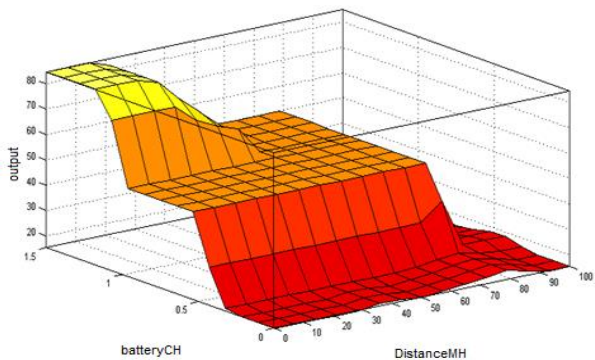

Figure 7. Surface graphic.

### E. The Fuzzy System Model

The Fuzzy Logic Model uses triangular and trapezoidal membership functions, Mamdani Inference System, Centroid Deffuzifier. During the CH selection process, to each input $(x_1, x_2, x_3)$, the system output is calculated as shown in (4).

$$y(x_1, x_2, x_3) = \frac{\sum_{l=1}^{27} \mu_{F_l^1}(x_1) \mu_{F_l^2}(x_2) \mu_{F_l^3}(x_3) c_{avg}^l}{\sum_{l=1}^{27} \mu_{F_l^1}(x_1) \mu_{F_l^2}(x_2) \mu_{F_l^3}(x_3)} \quad (4)$$

The selection of the leaders to participate of the data propagation process of the farthest CHs is defined by (5). Where $(x_1, x_2)$ are *BatteryCH* and *DistanceMH*.

$$y(x_1, x_2) = \frac{\sum_{l=1}^{9} \mu_{F_l^1}(x_1) \mu_{F_l^2}(x_2) c_{avg}^l}{\sum_{l=1}^{9} \mu_{F_l^1}(x_1) \mu_{F_l^2}(x_2)} \quad (5)$$

### F. The Network modeling

The network model has N sensors distributed in a N x M area. Three types of sensor nodes represent the heterogeneous network, normal, advanced and super sensor nodes, which have different starting energy levels. The calculation that determines the quantity of normal, advanced and super sensors is similar to the used by E-DEEC:

$$N.(1 - mf) \quad (6)$$

$$N. mf(1 - mp) \quad (7)$$

$$N. mf. mp \quad (8)$$

where $mf$ is the fraction of the total number of N sensors nodes and $mp$ the percentage of the total number of nodes that present more energy $e$ than the normal sensor nodes of the network. The calculation of the network total initial energy adopted is the same presented in [5].

$$E_{total} = N.(1 - mf).E_o + N.mf(1 - mp).(2).E_o + N.mf.mp.E_o(1 + e) = N.E_o(1 + mf(2 + mp.e)) \quad (9)$$

For the proposal with three heterogeneity levels, the network total energy is increased, considering the largest energy capacity of the advanced and super sensors. This difference is given by the factor $1 + mf(2 + mp.e)$.

### G. Network properties

The properties of the proposed network scenario are: (i) the distribution of the N network nodes is made in random way and does not have mobility; (ii) The nodes send their placement to the BS using GPS; (iii) The nodes dissipate energy to send the information; (iv) All nodes have the same processing and transmission capacity, the heterogeneity is applied to the energy levels, since some nodes have increased energy, differentiating them of the normal sensors; (v) The BS is fixed and its location is predefined in algorithm; (vi) The nodes always have data to transmit to the CH.

## V. SIMULATION RESULTS

The simulation is divided in rounds, at each round is obtained an output value based on the system input parameters and a new CH is elected to each *k cluster*. These values are updated to the input of the next round. Each round executes the phases described in section 4.

Each node is randomly distributed in a 100 m² area, where the classification of the number of normal, advanced and super nodes, with their respective energy levels is calculated using (5), (6) and (7), being $e = 1$, $mf = 1$ and $mp = 0.6$. The starting energy level for the normal nodes is 0.5J, for the advanced nodes is 1.0J and for the super nodes is 1.5J. The BS is previously defined with the coordinates x=5 and y=95. The radio model used is described in Table II.

TABLE II. RADIO MODEL PARAMETERS

| Parameters | Values |
|---|---|
| $E_{elec}$ | 50 $nJ/bit$ |
| $\varepsilon_{fs}$ | 10 $pJ/bit/m^2$ |
| $E_{amp}$ | 0.0013 $pJ/bit/m^4$ |
| $E_{DA}$ (Dissipated energy in data aggregation) | 5 $nJ/bit/signal$ |
| K-bit message (data) | 4000 $bits$ |
| K-bit message (info) | 100 $bits$ |
| $d_0$ (distance Threshold) | 87$m$ |

One hundred nodes are distributed, divided in *k* clusters. Each node sends 4000 Bits of message per round to the network cluster head. The data compression ratio is 5%. The simulation is made with 5000 rounds.

In this scenario is applied the metric FND (First Node Dies) to determine the network stability.

In the first phase of the simulation it is obtained the coordinates and energy level of each network node. The energy dissipated for the information sending of each node is calculated using (1). The initial energy of each node is decreased during this process. After the sending of the nodes coordinates, the algorithm K-means, defines a pattern based on the nodes coordinates, calculates the placement of each node and divides the nodes in *k* clusters. The algorithm also calculates the center of each cluster; this information is later used in the CH election process, described in sub-section *C*. The number of clusters used to the simulation is *k=5*.

Each node with its respective centrality value, proximity to the BS and energy level, will have a consequent, with pertinence degree *y* determining the chance of this node to become a CH. In the defuzzification process the node that have the highest crisp value is elected as optimal CH at current round. After, the phase 2 of the second stage is started, verifying if any node exceeds the communication threshold. If the threshold is exceeded the Fuzzy System determines the leader node more adequate for propagate the data of the farthest leader, considering its energy level and distance for the node that exceed the threshold.

The Fig. 8, shows the cluster division, within an area of 100 m², and the final CH election on round, being represented by '◊' Red, the cluster head elected by the system, and □ represents the BS. The circled nodes are the elected CHs that exceed the communication threshold with the BS.

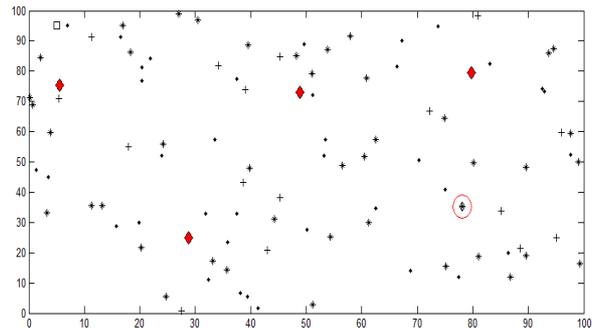

Figure 8. CHs selected by the fuzzy sistem on round 0.

TABLE I. TABLE III CLUSTER HEADS SELECTION RESULTS

| CH | Energy level | Cluster centrality | Number of nodes per cluster | Distance to BS |
|---|---|---|---|---|
| 1 | 1.4965 | 7.7574 | 21 | 19,6 m < $d_0$ |
| 2 | 1.4965 | 2.4817 | 21 | 49,0 m < $d_0$ |
| 3 | 1.4972 | 12.2433 | 17 | 76,3 m < $d_0$ |
| 4 | 1.4975 | 12.1966 | 15 | 94,5 m > $d_0$ |
| 5 | 1.4957 | 4.6183 | 26 | 73,9 m < $d_0$ |

The Table III presents the results of the leaders' selection on round 0. The energy values of the nodes shown already considers the energy dissipation generated for the information transmission to the collection point and also the dissipation that occurs during the association process to the elected CH. The CH 5 presented the higher dissipation due the quantity of associated nodes.

The CH 4, selected by the fuzzy system, exceeds the communication threshold. In a direct transmission to the collection point, this node would need to use the Multipath model, generating higher energy dissipation in the communication process, since the exponent of the path loss is $d^4$. The energy consumption generated in this kind of transmission would be approximately 2.3935 J, which drivers the node to premature inactivity and reducing the stability network period. However, with the multi-hop strate adopted, the CH communicates with the more adequate leader, keeping the radio Free space model. The Table IV shows the fuzzy system output, on round 0, for the selection of the adequate leader node for data propagation of the farthest CH. The higher crisp output value corresponds to the selected leader, CH 3.

TABLE II.　　　TABLE IV FUZZY SYSTEM OUTPUT

| CH | Energy level | Distance to the CH $> d_0$ | Fuzzy output |
|---|---|---|---|
| 1 | 1.4965 | 83.1566 | 18.2613 |
| 2 | 1.4965 | 47.8956 | 52.9779 |
| 3 | 1.4972 | 44.4752 | 57.3361 |
| 5 | 1.4957 | 50.3652 | 50.0000 |

The proposal of this paper was compared with the algorithms LEACH and E-DEEC. For performance evaluation it is used the network stability period and the useful lifetime. The choice of the algorithms for comparison is due principally by the use of local information, without considering placement criteria to the CH election. In addition of the leader choice method, the LEACH algorithm does not treat the energy discrepancies of the network nodes. Different from LEACH, the E-DEEC algorithm considers the heterogeneity of the nodes to the election of the CH. However, uses local information for leader election and inserts three levels of heterogeneity, similar to the proposal of this paper.

The Fig. 9 shows the quantity of active sensor nodes during the network useful lifetime. This measure reflects the total number of nodes that have not yet exhausted its energy in a given simulation period.

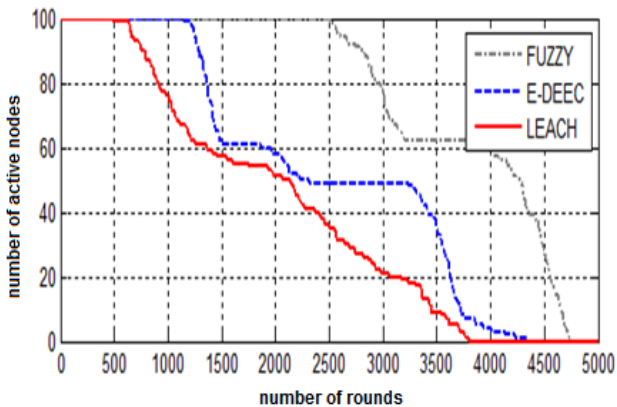

Figure 9. Number of active sensors during the simulation period.

The results clearly indicate that the insertion of heterogeneity levels, the fuzzy logic approach as selection tool, the use of multi-hops between the CHs and the BS centralized information, allow to elect more efficient leaders, increasing the stability period and the useful lifetime of the network.

The proposal presents better results when compared with LEACH and E-DEEC algorithms. the LEACH algorithm presented the lower stability period, occurring around 500 round. The E-DEEC algorithm presented better performance than LEACH, with instability period of the network starting around 1148 round. The proposal of this paper, as seen in Fig. 10, shows the better results over the compared algorithms, increasing the network stability period to approximately 2500 rounds, when occurs the first node inactivation due to lack of energy.

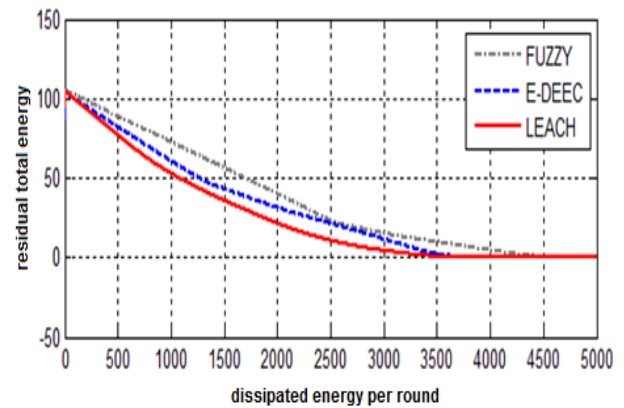

Figure 10. Total residual energy of LEACH, E-DEEC and FUZZY.

The Fig. 10 shows the network dissipated energy along 5000 simulation to each compared algorithm. The total energy for each network is 104.5J. The energy dissipation presents a linear decline along 2500 round for our proposal and 1500 rounds for E-DEEC, changing when the first node become inactive, breaking the stability period. Both Fuzzy and E-DEEC algorithm allow the treatment of the energetic discrepancies of each network node for the election of the CH, while the LEACH algorithm presents the higher energy dissipation per round, which is consequence of the problems already cited about the CH election way of the algorithm and lack of discriminatory treatment of the energetic discrepancies of the network nodes.

## VI. CONCLUSIONS

The results indicate that the proposal offers great benefits, allowing to select the most suitable cluster leaders at each round based on the values of the Fuzzy System defuzzification, as well as the use of fuzzy logic as a decision tool for implementing multiple hops between CHs, since it minimizes the energy dissipation of the selected CHs furthest from the collection point. The insertion of three levels of heterogeneity, corresponding to the normal, super and advanced sensors, contributes significantly to increasing the network stability period, since this insertion gives the network greater energy resources. However, the results indicate that if this difference is not taken into account when selecting the CH it does not influence significantly on increasing the stable period.

Another great advantage that contributes to the results obtained in this study is the use of a BS central. The BS does not have severe limitations of energy, processing and storage, as the nodes forming the network, thus, the BS has advantages over local processing of information on each node, this process, found in traditional algorithms for electing CHs, the sending of updates about the dissipated energy level information of nodes in each round. However, even with this update, the proposal still has improvements over other models presented. Another advantage of the BS central control occurs during the CH selection, having the role to inform the network about the leaders selected for each cluster, previously divided by k-means algorithm.

This process differs from the process found in algorithms that use local information to select their leader, the elected CH is responsible for sending broadcast messages to the network, generating energy dissipation at the moment of propagation. Finally, the work presented has the main contribution the election of the more efficient CH considering its location and energy level discrepancies, but also in adding new levels of heterogeneity, allowing thus increase the amount of network stability, ie the period that the network is fully functional, increasing considerably the lifetime of heterogeneous wireless sensor networks.